\documentclass[aps,pra,twocolumn,superscriptaddress,showpacs]{revtex4-1}
\usepackage{amsmath,amssymb,graphicx}
\usepackage{graphicx}

\begin{document}

\title{Optimizing optical Bragg scattering for single-photon frequency conversion}

\author{Simon Lefrancois}
\email[]{s.lefrancois@physics.usyd.edu.au}
\affiliation{Centre for Ultrahigh bandwidth Devices for Optical Systems (CUDOS), Institute of Photonics and Optical Science (IPOS), School of Physics, University of Sydney, NSW 2006, Australia}
\author{Alex S. Clark}
\affiliation{Centre for Ultrahigh bandwidth Devices for Optical Systems (CUDOS), Institute of Photonics and Optical Science (IPOS), School of Physics, University of Sydney, NSW 2006, Australia}
\affiliation{Centre for Cold Matter, Blackett Laboratory, Imperial College London, Prince Consort Road, London, SW7 2AZ, UK}
\author{Benjamin J. Eggleton}
\affiliation{Centre for Ultrahigh bandwidth Devices for Optical Systems (CUDOS), Institute of Photonics and Optical Science (IPOS), School of Physics, University of Sydney, NSW 2006, Australia}

\date{\today}

\begin{abstract}
We develop a systematic theory for optimising single-photon frequency conversion using optical Bragg scattering. The efficiency and phase-matching conditions for the desired Bragg scattering conversion as well as spurious scattering and modulation instability are identified. We find that third-order dispersion can suppress unwanted processes, while  dispersion above the fourth order limits the maximum conversion efficiency. We apply the optimisation conditions to frequency conversion in highly nonlinear fiber, silicon nitride waveguides and silicon nanowires. Efficient conversion is confirmed using full numerical simulations. These design rules will assist the development of efficient quantum frequency conversion between multicolour single photon sources for integration in complex quantum networks.
\end{abstract}

%\keywords{}

\maketitle

\section{Introduction}

Quantum information and computing using single-photons is a promising way of realizing qubits which are stable against the environment yet can be manipulated and detected readily~\cite{OBrien2007}. A number of current schemes for developing quantum photonic technology such as quantum gates~\cite{Politi:08}, algorithms~\cite{Politi:09}, and metrology~\cite{bell:13} require the use of indistinguishable photons, as they need high visibility Hong-Ou-Mandel (HOM) quantum interference~\cite{hong:1987}. Although quantum interference has been observed between disparate sources~\cite{McMillan2013}, this required extremely careful engineering of the spectral states.  To avoid this in the future and have an agile quantum network, quantum frequency conversion will be required to bring together photons from separate nodes.  A method to achieve this conversion is to use optical nonlinearities to convert between states. Parametric wave mixing in the single photon regime has been proposed using second and third-order nonlinear media~\cite{Kumar:90,Albota:04,McKinstrie:05,takesue:2008}. 

A promising quantum frequency conversion technique is Bragg scattering (BS) four-wave mixing (FWM) between single photons and two strong classical pumps in third-order nonlinear optical media~\cite{inoue:1994,McKinstrie:05}. This noiseless conversion process conserves quantum state information and can enable quantum interference between photons at different frequencies~\cite{McGuinness:11}. However, a drawback of Bragg scattering in third-order media is competition with other parametric processes, such as modulation instability (MI) and reverse Bragg scattering~\cite{mckinstrie:02}, which can be phase-matched simultaneously if proper care is not taken. This largely explains why several demonstrations of single photon BS quantum frequency conversion suffered from low efficiency~\cite{mcguiness:2010,Agha:12}. Recently, highly efficient single-photon frequency conversion was demonstrated in a highly-nonlinear fiber~\cite{Clark:13}. So far, the optimisation of the frequency conversion process has been done mostly on an ad hoc basis for a given experiment. It would thus be valuable to develop a systematic process for optimizing the desired frequency conversion process while suppressing unwanted parametric mixing. This will be particularly critical to achieve conversion efficiencies above 50\%, a requirement for unity visibility multicolor quantum interference between photons of different frequency~\cite{Sharping:06,Hunault:10,Xiong:11}.

In this article, we develop a systematic theory for optimising Bragg scattering frequency conversion for maximum efficiency in the single-photon regime. The desired and spurious parametric processes are identified. The efficiency and phase-matching conditions for backward and forward Bragg scattering and spurious modulation instability are clearly described using analytical theory. We find that large third-order dispersion can be used to suppress spurious processes while leaving the desired frequency conversion unaffected. Meanwhile, higher-order dispersion from the fourth order and up limits the maximum frequency conversion efficiency even before spurious processes come into play. Our general theory is applied to three candidate media for single photon Bragg scattering: highly nonlinear fiber, integrated silicon nitride waveguides, and silicon nanowires. We find conditions of optimal operation for properly selected dispersion profiles, and confirm this with full numerical simulations of the simultaneous parametric processes. This work will provide a guide to design efficient and low-noise quantum state translation devices for integration into quantum information circuits. The conclusions also apply to classical optical Bragg Scattering systems where spurious effects need to be avoided.

\section{Frequency conversion theory}

\begin{figure}[htp]
\includegraphics{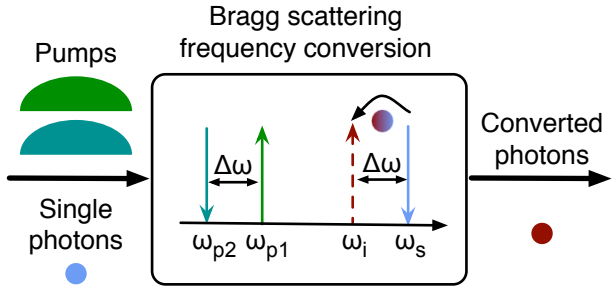}
\caption{ (Color online) Single photon frequency conversion using optical Bragg scattering. Two classical pumps at frequencies $\omega_{p_1}$ and $\omega_{p_2}$ convert single photons from $\omega_{s}$ to $\omega_i$.}
\label{fig:fwm_concept}
\end{figure}

The basic concept of single-photon frequency translation using optical Bragg scattering is illustrated in Fig.~\ref{fig:fwm_concept}. Single photons at the signal frequency $\omega_s$ interact with two strong classical pumps in a $\chi^{(3)}$ nonlinear medium and are scattered to a new idler frequency $\omega_i$. Energy conservation dictates that the BS frequency shift is $\Delta\omega=\omega_{p1}-\omega_{p2} = \omega_{s}-\omega_{i}$. The process conserves photon number in the signal and idler channels, allowing for noiseless frequency conversion preserving quantum state information~\cite{McGuinness:11}.

\begin{figure}[htp]
\includegraphics{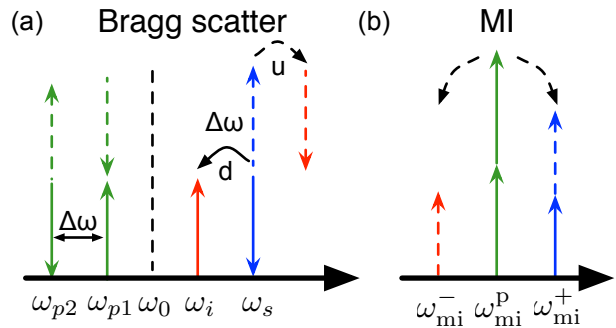}
\caption{ (Color online) (a) Bragg scattering frequency conversion. (b) Modulation instability amplification. The desired frequency conversion is in solid lines, the spurious processes are dashed.}
\label{fig:fwm_proc}
\end{figure}

In addition to Bragg scattering, four-wave mixing processes also include modulation instability (also known as degenerate or single-pump FWM) and phase conjugation~\cite{mckinstrie:02}. The processes can compete with the desired frequency translation process. The two main spurious processes of interest, reverse Bragg scattering and parametric amplification through modulation instability, are illustrated in Fig.~\ref{fig:fwm_proc}. Here we will optimise for the BS process bringing the idler closer to the pumps, labeled down in Fig.~\ref{fig:fwm_proc}(a). This choice is arbitrary and the conclusions drawn also apply for the reverse scattering process labeled up. If the phase matching conditions are not optimized, both up- and down-conversion can be phase-matched simultaneously, reducing the efficiency of the desired process. Additionally, modulation instability can cause the generation of secondary pump wavelengths and depletion of the main pumps, as well as couple vacuum noise into the quantum state wavelengths. For MI, energy conservation requires that $2\omega_{\text{mi}}^p=\omega_{\text{mi}}^++\omega_{\text{mi}}^-$, as defined in Fig.~\ref{fig:fwm_proc}(b).

\subsection{Bragg scattering}

The Bragg scattering process on single photons can be described in terms of the classical pump fields and the quantum mechanical single photon signal and idler fields. We assume that the pumps each have equal powers $P_{p1}=P_{p2}=P_0$. The spatial Hamiltonian is then~\cite{McKinstrie:05}:
\begin{equation}
H/\hbar = \delta_{\text{bs}}(a_s^\dag a_s-a_i^\dag a_i) + \kappa a_s^\dag a_i + \kappa^* a_i^\dag a_s.
\label{eq:hamil}
\end{equation}
The photon creation and annihilation operators are $a^\dag_j$ and $a_j$ respectively. The linear wave-vector mismatch is $\delta_{\text{bs}}$ and the pump-induced coupling is $\kappa = 2\gamma P_0$ with $\gamma$ the nonlinear Kerr coefficient. The evolution of the creation/annihilation operators is then governed by the spatial Heisenberg equation of motion:
\begin{equation}
i\hbar\frac{da_j}{dz} = [H,a_j].
\label{eq:heisen}
\end{equation}

We can solve Eqs.~\ref{eq:hamil}-\ref{eq:heisen} to get the output form of the operators after propagation~\cite{McKinstrie:05}:
\begin{align}
a_s(z) &= \mu(z)a_s(0) + \nu(z)a_i(0),\label{eq:asz}\\
a_i(z) &= -\nu^*(z)a_s(0) + \mu^*(z)a_i(0),\label{eq:aiz}
\end{align}
with the transfer functions:
\begin{align}
\mu(z) &= \cos{(k_{\text{bs}}z)} + i\delta_{\text{bs}}\sin{(k_{\text{bs}}z)}/k_{\text{bs}},\label{eq:mu}\\
\nu(z) &= i\kappa\sin{(k_{\text{bs}}z)}/k_{\text{bs}}.\label{eq:nu}
\end{align}
We have used the Bragg scattering wavevector $k_{\text{bs}}$:
\begin{equation}
k_{\text{bs}} = \sqrt{\delta_{\text{bs}}^2+\kappa^2}.
\label{eq:kbs}
\end{equation}

Consider the case where the input photon number state $|n_s,n_i\rangle$ consists of a single photon at the signal frequency, described as $|\psi(0)\rangle=|1,0\rangle=a_s^\dag(0)|0,0\rangle$. The propagating state can be found using the unitary operator $U=\exp{(-iHz/\hbar)}$:
\begin{align}
|\psi(z)\rangle &=U|\psi(0)\rangle=Ua_s^\dag(0)|0,0\rangle\nonumber\\
& = (Ua_s^\dag(0)U^\dag) U|0,0\rangle = a_s^\dag(z)|0,0\rangle,
\end{align}
where we have used the invariance of the vacuum state under propagation. Using Eq.~\ref{eq:asz}, we get:
\begin{equation}
|\psi(z)\rangle= \mu^*(z)|1,0\rangle + \nu^*(z)|0,1\rangle.
\end{equation}
Thus, the Bragg scattering process propagates the input single photon into signal and idler photon states according to the transfer functions $\mu(z)$ and $\nu(z)$. Photon numbers are conserved since $|\mu|^2+|\nu^2|=1$.

We can now define a Bragg scattering conversion efficiency $\eta_{\text{bs}} = |\langle 0,1 |\psi\rangle|^2$, which corresponds to the probably of finding the photon in the idler state:
\begin{align}
\eta_\text{bs}(z) &= |\nu(z)|^2 =  \bigg(\frac{2\gamma P_0}{k_{\text{bs}}}\bigg)^2\sin^2{(k_{\text{bs}}z)}.\label{eq:idlerconv}
\end{align}
Similarly, we can define a signal transmittance $\tau_\text{bs}= |\langle 1,0 |\psi\rangle |^2$ and show from Eq.~\ref{eq:mu}-\ref{eq:nu}:
\begin{align}
\tau_\text{bs}(z) = |\mu(z)|^2 = 1 - \eta_\text{bs}(z).
\end{align}
These results are consistent with the classical treatment of Bragg scattering~\cite{mckinstrie:02}. This is expected because at the quantum mechanical level, the wave-mixing process operate linearly on the quantum states, as can be seen from Eq.~\ref{eq:hamil}.

The dispersive phase-mismatch $\delta_{\text{bs}}= (\beta_i-\beta_s+\beta_{p1}-\beta_{p2})/2$ can be expressed in terms of the dispersion $\beta_n(\omega_0) = \partial^n\beta(\omega)/\partial\omega^n\big|_{\omega0}$, where $\beta(\omega)$ is the waveguide propagation constant:
\begin{align}
{\delta_{\text{bs}}}=\sum_n\frac{\beta_{2n}(\omega_0)}{(2n)!}\big[ (\omega_{p2}-\omega_0)^{2n} - (\omega_{p1}-\omega_0)^{2n} \big],
\label{eq:dbbs}
\end{align}
with $\omega_0$ the centre frequency of the BS process. For the desired down scattering we have $\omega_0^{d}=(\omega_{p2}+\omega_s)/2$ while for spurious up scattering we have $\omega_0^{u}=(\omega_{p1}+\omega_s)/2$. The maximum conversion efficiency for the desired Bragg scattering down-conversion is then:
\begin{align}
\eta_{\text{bs}}^d = \bigg(\frac{2\gamma P_0}{k_\text{bs}}\bigg)^2 = \frac{1}{1+{\delta_{\text{bs}}^-}^2/(2\gamma P_0)^2}.
\label{eq:cebs}
\end{align}
If the BS centre frequency $\omega_0$ is chosen to be a zero-dispersion wavelength and we neglect dispersion beyond $\beta_2$, we get $k_{\text{bs}}=2\gamma P_0$ and $\eta^d_{\text{bs}}=1$ for all pump wavelengths. For the rest of the paper, we will work under this regime to maximise the frequency conversion bandwidth. If higher-order dispersion is strong enough it will nevertheless limit the maximum pump spacing as we will see below.

From Eq.~\ref{eq:idlerconv}, the maximum conversion into the idler is obtained when $k_{\text{bs}}z=\pi/2$. Assuming perfect phase-matching $\delta_{\text{bs}} = 0$ along a waveguide of length $L$, this corresponds to a single pump power:
\begin{equation}
P_0^{\text{max}} = \frac{\pi}{4\gamma L}
\label{eq:pmax}
\end{equation} 

The first process to compete with the desired frequency conversion is the opposite BS process. In the case where up-conversion is the unwanted BS process, Eq.~\ref{eq:dbbs} still applies, but with the dispersion $\beta_n(\omega_0^u)$. This unwanted BS can be avoided if the phase-mismatch $\delta_{\text{bs}}^u$ is large enough. To first order, this is dominated by the group-velocity dispersion $\beta_2(\omega_0^u) = \beta_2(\omega_0^d)+\beta_3(\omega_0^d)(\omega_0^u-\omega_0^d) = \beta_2(\omega_0^d)-\beta_3(\omega_0^d)\Delta\omega/2$. Thus, spurious frequency conversion can be minimized by moving the pumps away from the center frequency and increasing the pump spacing, although higher-order dispersion may reduce conversion efficiency for the desired conversion at large spacings. Interestingly, the third-order dispersion (TOD) $\beta_3$ can also mitigate spurious conversion without affecting the desired frequency conversion. This is discussed in more details in the context of modulation instability below and in Fig.~\ref{fig:hnlf_dispdemo}, where TOD has a similar effect.

\subsection{Modulation instability}

In addition to spurious Bragg scattering, spurious parametric amplification from modulation instability also competes with frequency translation by depleting the pumps and forming secondary pump wavelengths. It can also couple noise into the quantum channels. To understand how to minimize this, we first consider the classical parametric amplification gain $G_{\text{mi}}$~\cite{mckinstrie:02}:
\begin{align}
G_{\text{mi}}(z) &= 1 + \bigg(\frac{\gamma P_0}{g_{\text{mi}}}\bigg)^2\sinh^2{(g_{\text{mi}}z)},
\label{eq:gainbs}
\end{align}
where the parametric gain coefficient $g_{\text{mi}}$ is:
\begin{align}
g_{\text{mi}} = \sqrt{(\gamma P_0)^2-(\delta_{\text{mi}}+\gamma P_0)^2}.
\label{eq:gmi}
\end{align}
The linear phase-mismatch for modulation instability is $\delta_{\text{mi}}= (\beta_s+\beta_i-2\beta_{p})/2$. In terms of the dispersion we have:
\begin{align}
{\delta_{\text{mi}}}=\sum_n\frac{\beta_{2n}(\omega_p)}{(2n)!} (\omega_{\text{mi}}^+-\omega^p_{\text{mi}})^{2n}.
\label{eq:dbmi}
\end{align}
Spurious modulation instability can occur between any pair of waves involved in the BS process. Generally, the dominant contribution will be MI between the BS pumps, given their high powers. This corresponds to $\omega^p_{\text{mi}}=\omega_{p1,p2}$ and $\omega^+_{\text{mi}}=\omega_{p2,p1}$. This describes the generation of secondary pumps and depletion of the pumps through cascaded FWM. Another process where $\omega^\pm_{\text{mi}}=\omega_{i,s}$ represents spurious parametric amplification of the quantum states, which can couple shot noise into the quantum channels.

\begin{figure}[htp]
\includegraphics{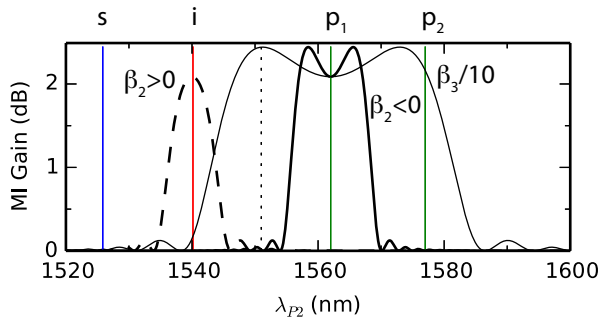}
\caption{ (Color online) Effects of dispersion on spurious modulation instability: optimised process (thick solid), low third-order dispersion (thin solid, $\beta_3$/10) and normal dispersion (dashed). Parameters are those of the fiber in Sec.~\ref{sec:hnlf} with $\lambda_{p1} = 1562$~nm and $\lambda_{p2} = 1577$~nm. The pump, signal, idler (vertical solid) and centre wavelength (dotted) for the desired frequency conversion are marked for reference.}
\label{fig:hnlf_dispdemo}
\end{figure}

Both processes can be minimized similarly to the BS process described above, as shown in Fig.~\ref{fig:hnlf_dispdemo}. Higher third-order dispersion reduces the bandwidth of the spurious processes (thin and thick solid lines). In addition, the MI bandwidth can also be narrowed by placing the pumps in the normal dispersion regime (dashed lines), where dispersion cannot cancel the nonlinear phase-mismatch as can be seen from Eq.~\ref{eq:gmi}. However, if the normal dispersion regime is on the short wavelength side, it may be preferable to place the pumps on the long wavelength side to reduce contamination of the signal by spontaneous Raman scattering~\cite{Clark:13}.

\subsection{Loss and numerical modelling}

When a propagation loss $\alpha$ is present, the BS wavenumber and MI parametric gain vary along the waveguide. It is possible to derive general expressions for $k$ and $g$ in terms of hypergeometric functions~\cite{marhic2008book}. However, here we want to extract simple design guidelines and want to avoid excessive loss, so these are needlessly complicated. It can be shown that an approximation in the presence of loss is obtained by replacing $L$ with an effective length $L_{\text{eff}}$ in all calculations above (see Appendix ~\ref{sec:leff}):
\begin{equation}
L_{\text{eff}} = \frac{1-e^{-\alpha L}}{\alpha}.
\label{eq:leff}
\end{equation}
This implies normalizing out the loss $e^{-\alpha L}$, which can be lumped into the total system loss. This effective length treatment captures the effects of loss on the classical pumps and the associated wave-mixing processes. For the quantum single photon states, losses couple vacuum noise into the system and can yield mixed states~\cite{Drummond:01}. Since the focus of the present work is to optimise frequency conversion and suppress spurious processes, we defer a full quantum mechanical treatment of loss to future work.

The BS conversion efficiency Eq.~\ref{eq:cebs} and spurious MI gain Eq.~\ref{eq:gainbs} above were derived using the four-wave undepleted pump approximation, which cannot account for gain saturation effects or secondary mixing between spurious products. To model the full FWM interactions, we numerically solve the nonlinear Schr\"odinger equation using the split-step Fourier method~\cite{agrawal2013book}:
\begin{multline}
\frac{\partial A}{\partial z} = -\frac{\alpha}{2}A + \sum_{n=2}^\infty i^{n+1}\frac{\beta_n}{n!}\frac{\partial^n A}{\partial t^n}  \\
+i\gamma\bigg( 1+\frac{i}{\omega_c}\frac{\partial}{\partial t} \bigg)\big(|A|^2A\big).
\label{eq:nlse}
\end{multline}
The field envelope $A(z,t)$ is defined such that the power $P=|A|^2$ and includes all waves involved in the FWM processes. The nonlinear derivative dependant on the carrier frequency $\omega_c$ ensures conservation of energy in the frequency conversion process.

Raman scattering between the single photons and pumps can also drain away the quantum channels and couple in spontaneous noise~\cite{Clark:13,Collins:12}. This can be modelled by adding a delayed nonlinear response function~\cite{BLOW:1989aa,Hollenbeck:02} and spontaneous scattering term~\cite{Lin:2007pra} to Eq.~\ref{eq:nlse}. Here we will neglect this effect and focus on phase-matching of parametric processes, since the Raman response function is not always available.

The theory and simulations used here assume continuous wave fields. The general conclusions on four-wave mixing phase-matching will also apply to pulsed fields provided care is taken with the pulse lengths~\cite{McGuinness:11}.

\section{Highly nonlinear fiber}
\label{sec:hnlf}

We first demonstrate optimization of the frequency conversion in a highly-nonlinear fiber (HNLF), similar to previous experiments~\cite{Clark:13}. The fiber length is $L=0.75$~km and we consider loss negligible over this distance. The nonlinearity is $\gamma=21$~(W km)$^{-1}$. This corresponds to an estimated peak conversion power of 50~mW per pump from Eq.~\ref{eq:pmax}. The higher-order dispersion around the zero-dispersion wavelength of 1551~nm (where $\beta_2=0$) is $\beta_3 = 0.032$~ps$^3$/km and $\beta_4 = -5.4\cdot 10^{-5}$~ps$^4$/km. 

\begin{figure}[htp]
\includegraphics{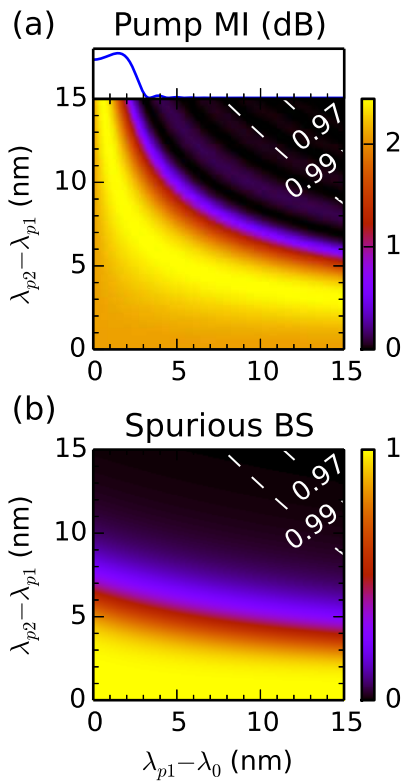}
\caption{ (Color online) Four-wave mixing in an HNLF with $P_0=50$~mW and centre wavelength $\lambda_0=1551$~nm (also the zero-dispersion point). (a) Modulation instability gain from pump 1 with idler (top, normalized) and with pump 2 (bottom). (b) Spurious Bragg scattering conversion efficiency and contours of desired BS efficiency (dotted white).}
\label{fig:hnlf2d}
\end{figure}

The strength of the spurious frequency mixing processes are shown in Fig.~\ref{fig:hnlf2d}. The frequency conversion process is centered at the zero-dispersion wavelength of the HNLF to maximise its bandwidth, as discussed before. Spurious modulation instability gain can be avoided for pump spacings greater than 5-10~nm, while frequency conversion remains highly efficient over this range, as shown by the dashed contour lines. Spurious Bragg scattering can also be avoided for spacings greater than 5~nm.

\begin{figure}[htp]
\includegraphics{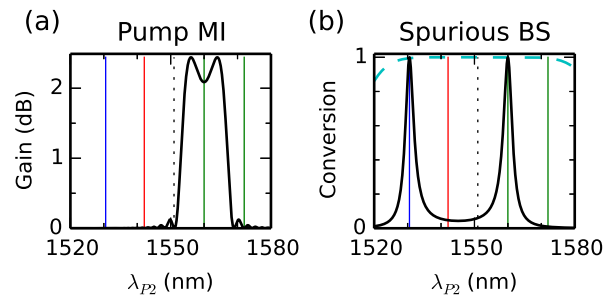}
\caption{ (Color online) (a) Modulation instability gain of pump 1 in an HNLF. (b) Conversion efficiency of spurious (solid) and desired (dashed) Bragg scattering in an HNLF. The pump power is $P_0=50$~mW. The pump, signal, idler (solid) and centre wavelength (dotted) are marked for reference.}
\label{fig:hnlfpm}
\end{figure}

Guided by the two-dimensional optimization in Fig.~\ref{fig:hnlf2d}, we choose pump wavelengths $\lambda_{p1}=1560$~nm and $\lambda_{p2}=1572$~nm where dispersion is anomalous ($\beta_2<0$), closely matching previous experiments. The corresponding signal and idler wavelengths are $\lambda_{s}=1530.6$~nm and $\lambda_{i}=1542.1$~nm. The resulting spurious processes spectra are shown in Fig.~\ref{fig:hnlfpm}. As expected, unwanted frequency conversion is minimal.

\begin{figure}[htp]
\includegraphics{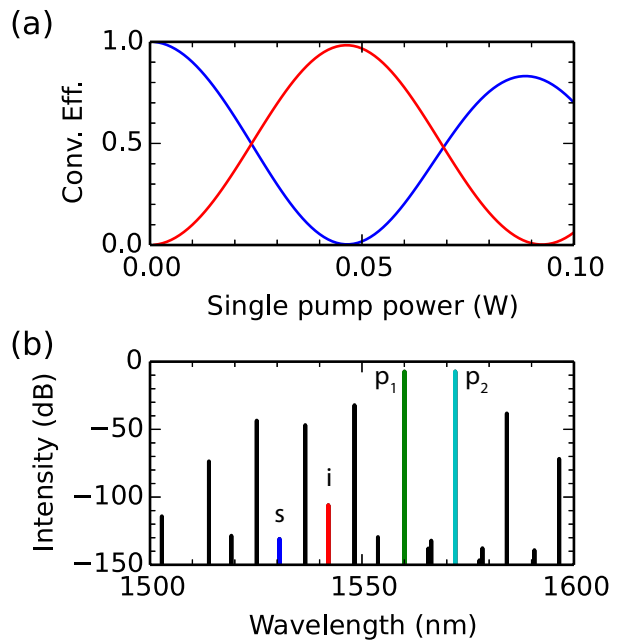}
\caption{ (Color online) (a) Simulated evolution of input signal (1530.6~nm, blue) and converted idler (1542.1~nm, red) in an HNLF. (b) Output spectrum for input pump power $P_0 = 46.5$~mW.}
\label{fig:hnlfsim}
\end{figure}

The full numerical simulations of the frequency conversion process in HNLF are shown in Fig.~\ref{fig:hnlfsim}. The input signal power $P_s^0 = 6.4$~pW corresponds to the average power for a single photon rate of 50~MHz. The conversion efficiency is not affected by the exact signal power since the mixing processes are all linear in the signal field. Figure~\ref{fig:hnlfsim}(a) shows that simulations predict 98\% frequency conversion, as observed in experiments~\cite{Clark:13}. This gives confidence in the reliability of the optimisation process presented here. The maximum conversion occurs for pump powers of 46.5~mW, slightly lower than the analytic prediction. This is probably due to the effects of slight phase-mismatch and spurious conversion which make the powers and BS wavevector vary over propagation, but is a small correction. Figure~\ref{fig:hnlfsim}(b) shows the output spectrum at maximum conversion. Spurious MI and BS are visible, but at least 20~dB below the desired signals, and well separated spectrally, indicating proper optimisation of the frequency conversion. 

\section{Silicon nitride waveguide}

\begin{figure}[htp]
\includegraphics{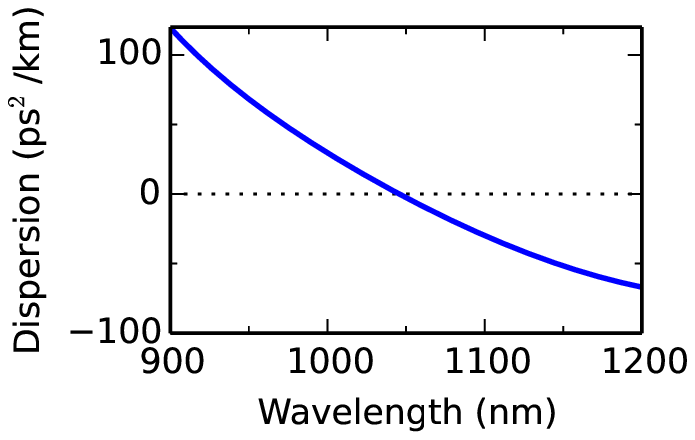}
\caption{ (Color online) Simulated dispersion of a $550\times1000$~nm$^2$ silicon nitride waveguide in the TE mode from Ref.~\cite{Agha:12}.}
\label{fig:sin_disp}
\end{figure}

We now optimise frequency conversion in an on-chip silicon nitride waveguide experimentally described in Ref.~\cite{Agha:12}. Nanophotonics waveguides tend to display considerable higher-order dispersion, limiting frequency conversion efficiency. The dispersion profile can be optimised by varying the waveguide dimensions. We choose dimensions of $550\times1000$~nm$^2$ and operate around the 1~$\mu$m range in the TE mode where TOD is strong enough while higher-order dispersion is moderate, as shown in Fig.~\ref{fig:sin_disp}. We use a length of $L=2.5$~cm, which corresponds to an effective length $L_{\text{eff}}=1.9$~cm for a propagation loss $\alpha=1$~dB/cm. We use a nonlinearity of $\gamma=7.6$~(W m)$^{-1}$ corresponding to $P_0^{\text{max}}=5.4$~W, which can be reached using picosecond pulses~\cite{Agha:12}. The dispersion coefficients around the zero-dispersion wavelength at 1045.6~nm are $\beta_2 = -9.6\cdot10^{-3}$~ps$^2$/km, $\beta_3 = 0.35$~ps$^3$/km, $\beta_4 = 2.6\cdot10^{-4}$~ps$^4$/km, $\beta_5 = -1.2\cdot10^{-6}$~ps$^5$/km and $\beta_6 = 3.8\cdot10^{-8}$~ps$^6$/km. 

\begin{figure}[htp]
\includegraphics{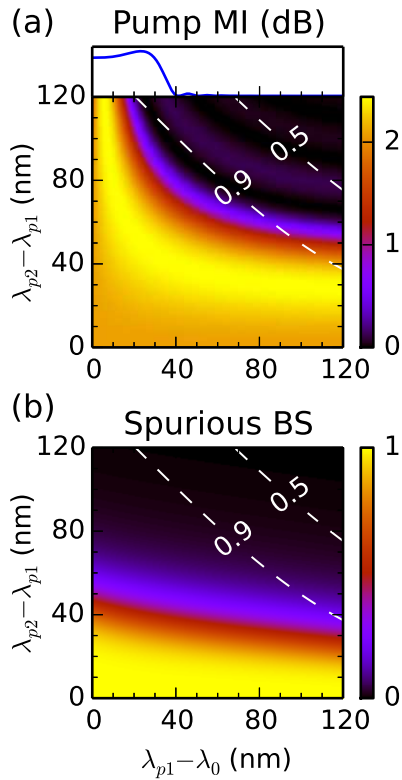}
\caption{(Color online) Four-wave mixing in a silicon nitride waveguide with $P_0=5.4$~W at $\lambda_0=1045.6$~nm. (a) Modulation instability gain from pump 1 with idler (top, normalized) and with pump 2 (bottom). (b) Spurious Bragg scattering conversion efficiency and contours of desired BS efficiency (dotted white). The global loss of 2.5~dB has been normalized out.}\label{fig:sin2d}
\end{figure}

The spurious processes are mapped out in Fig.~\ref{fig:sin2d} around the zero-dispersion wavelength $\lambda_0 = 1045.6$~nm. Note that we have normalized out the total linear loss of 2.5~dB as discussed above. Pump spacings on the order of 50~nm are required to avoid MI and reverse BS. This is because the short lengths and high powers needed in nanophotonic waveguides yield small dispersive phase-mismatches and strong FWM conversion. The frequency conversion efficiency also drops off more quickly due to the larger higher order dispersion. Nevertheless, a window with low spurious conversion and conversion efficiency above 90\% is available.

\begin{figure}[htp]
\includegraphics{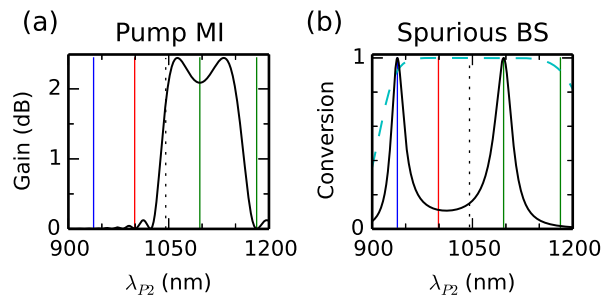}
\caption{ (Color online) (a) Modulation instability gain of pump 1 in a silicon nitride waveguide of at anomalous (solid) and normal (dotted) dispersion. (b) Conversion efficiency of spurious (solid) and desired (dashed) Bragg scattering in an HNLF. The pump power is $P_0=5.4$~W. The pump, signal, idler (solid) and centre wavelength (dotted) are marked for reference. The global loss of 2.5~dB has been normalized out.}
\label{fig:sin_pm}
\end{figure}

Following Fig.~\ref{fig:sin2d}, we choose pump wavelengths $\lambda_{p1}=1096.6$~nm and $\lambda_{p2}=1181.6$~nm, corresponding to signal and idler wavelengths $\lambda_{s}=937.7$~nm and $\lambda_{i}=999.1$~nm. The pump range can be accessed using Yb-doped fiber lasers optimised for long-wavelength operation~\cite{pask:1995}. The resulting parametric gain spectra are shown in Fig.~\ref{fig:sin_pm}, showing that spurious processes are indeed avoided.

\begin{figure}[htp]
\includegraphics{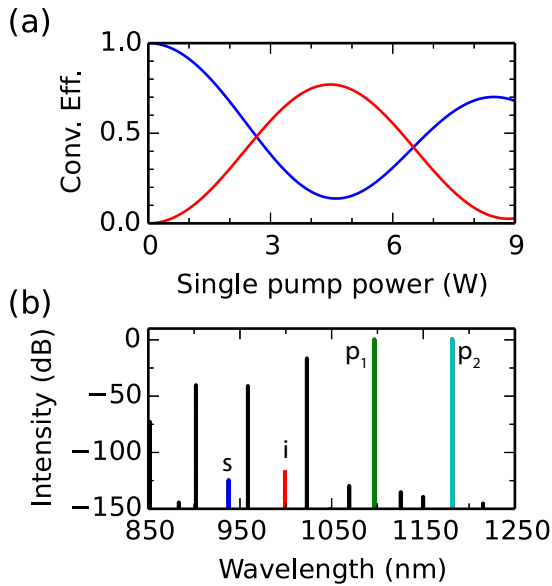}
\caption{ (Color online) (a) Simulated evolution of input signal (937.7~nm, blue) and converted idler (999.1 ~nm, red) in a silicon nitride waveguide. (b) Output spectrum for input pump power $P_0 = 4.5$~W. The global loss of 2.5~dB has been normalized out.}
\label{fig:sin_sim}
\end{figure}

Full numerical simulation for the silicon nitride waveguide are shown in Fig.~\ref{fig:sin_sim}. A peak conversion of 75\% is achieved for a single pump power of 4.5~W. This is again lower than the analytic prediction, but still close. The conversion efficiency is lower than the expected 90\%. This is likely due to the full effects of loss, which continuously shift the phase-matching condition, and the presence of multiple parametric mixing products visible in Fig.~\ref{fig:sin_sim}(b). Nevertheless, this shows that frequency conversion with efficiency above 50\% is possible in properly optimised waveguides, which can enable Hong-Ou-Mandel (HOM) interference between frequency non-degenerate photon pairs generated by spontaneous parametric processes.

\section{Silicon waveguide}

Finally, we model frequency conversion in an on-chip silicon nanowire. Besides CMOS integration potential~\cite{Lipson:2005}, crystalline silicon displays a narrow Raman resonance which can be avoided, as opposed to broadband Raman in glasses which can contaminate single photon signals. A drawback of silicon is multiphoton absorption, which can limit the effective nonlinearity~\cite{Yin:07}. However, as we will show below, for the powers used here two-photon absorption is low.

\begin{figure}[htp]
\includegraphics{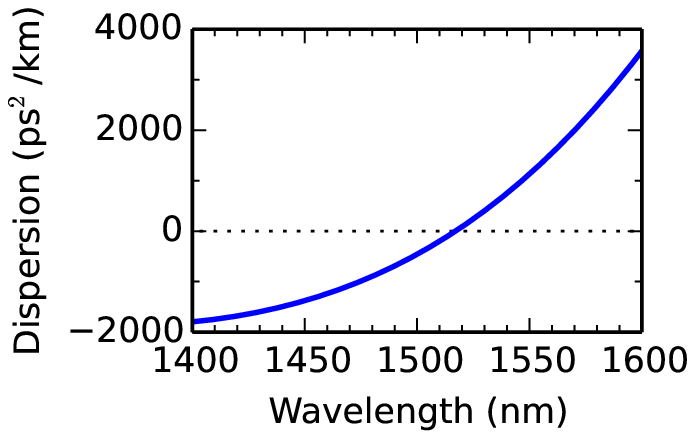}
\caption{ (Color online) Simulated dispersion of a $300\times300$~nm$^2$ silicon waveguide in the TM mode from Ref.~\cite{Turner:06}.}
\label{fig:silicon_disp}
\end{figure}

Dispersion engineering of such waveguides is well documented~\cite{Turner:06}. Dimensions of $300\times300$~nm$^2$ are chosen to maximise the third-order dispersion around 1.5~$\mu$m, as shown in Fig.~\ref{fig:silicon_disp}. We use a length of $L=1.0$~cm, which corresponds to an effective length $L_{\text{eff}}=0.76$~cm for a propagation loss $\alpha=2.5$~dB/cm. The nonlinearity of $\gamma=350$~(W m)$^{-1}$ corresponds to $P_0^{\text{max}}=0.295$~W. The dispersion coefficients around 1516~nm are $\beta_2 = -31$~ps$^2$/km, $\beta_3 = -35$~ps$^3$/km, $\beta_4 = 0.50$~ps$^4$/km, $\beta_5 = -4.9\cdot10^{-3}$~ps$^5$/km and $\beta_6 = 3.6\cdot10^{-5}$~ps$^6$/km. As for two-photon absorption, using a typical value of $\alpha_{TPA} = 10\cdot10^{-12}$~m/W and an effective area $A_{\text{eff}}=0.1$~$\mu$m$^2$~\cite{bristow:07}, the TPA length scale is $(\alpha_{TPA}P_0/A_{\text{eff}})^{-1} = 30$~cm. This is much longer than the waveguide used, thus two-photon losses can be neglected.

\begin{figure}[htp]
\includegraphics{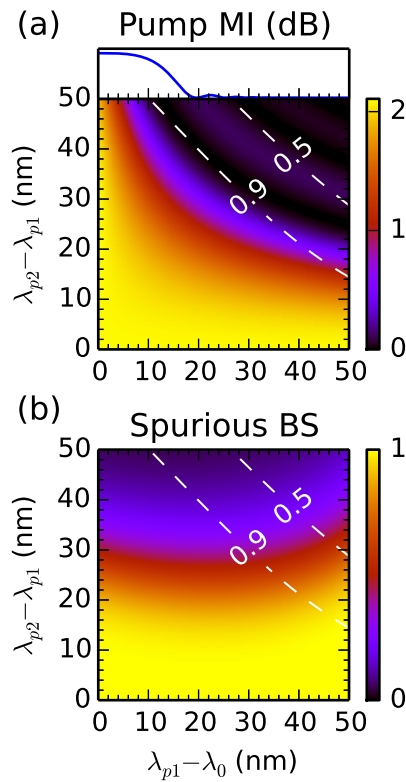}
\caption{(Color online) Four-wave mixing in a silicon waveguide with $P_0=295$~mW at $\lambda_0=1516$~nm. (a) Modulation instability gain from pump 1 with idler (top, normalized) and with pump 2 (bottom). (b) Spurious Bragg scattering conversion efficiency and contours of desired BS efficiency (dotted white). The global loss of 2.5~dB has been normalised out.}
\label{fig:silicon_2d}
\end{figure}

\begin{figure}[htp]
\includegraphics{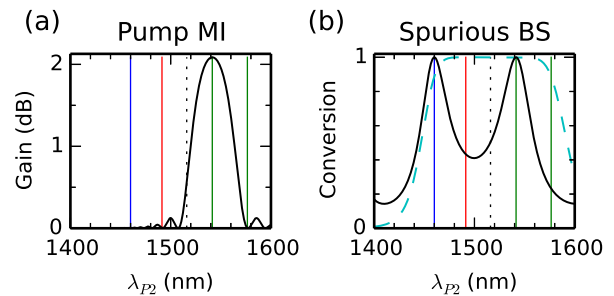}
\caption{ (Color online) (a) Modulation instability gain of pump 1 in a silicon waveguide at anomalous (solid) and normal (dotted) dispersion. (b) Conversion efficiency of spurious (solid) and desired (dashed) Bragg scattering in an HNLF. The pump power is $P_0=295$~mW. The pump, signal, idler (solid) and centre wavelength (dotted) are marked for reference. The global loss of 2.5~dB has been normalized out.}
\label{fig:silicon_pm}
\end{figure}

The spurious parametric processes in the silicon nanowire are modelled around 1516~nm in Fig.~\ref{fig:silicon_2d}. The propagation loss of 2.5~dB has been normalised out. Modulation instability can be avoided with conversion efficiencies around 90\%. However, spurious Bragg scattering is harder to avoid while maintaining both efficient conversion and low MI gain, as seen in Fig.~\ref{fig:silicon_pm}(b). This is due to the relatively large fourth-order dispersion, which limits the pump spacings range where frequency translation is efficient. Indeed, for the silicon nanowire we have $\beta_4/\beta_3 \sim 10$~fs, while for the HNLF and the silicon nitride waveguide the ratio is on the order of 1~fs.

We chose to model pump wavelengths $\lambda_{p1}=1541.5$~nm and $\lambda_{p2}=1576.5$~nm, corresponding to signal and idler wavelengths $\lambda_{s}=1460.0$~nm and $\lambda_{i}=1491.3$~nm. The resulting gain spectra are shown in Fig.~\ref{fig:silicon_pm}. As expected, modulation instability is well mitigated, but spurious Bragg scattering efficiencies around 20\% are expected.

\begin{figure}[htp]
\includegraphics{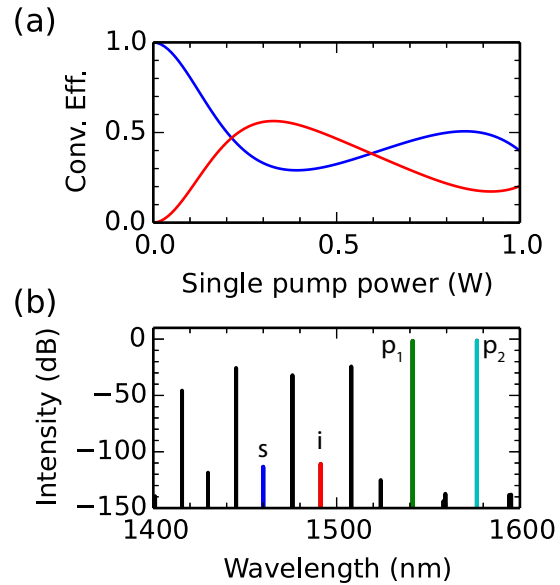}
\caption{ (Color online) (a) Simulated evolution of input signal (1459.9~nm, blue) and converted idler (1491.3~nm, red) in a silicon waveguide. (b) Output spectrum for input pump power $P_0 = 320$~mW. The global loss of 2.5~dB has been normalized out.}
\label{fig:silicon_sim}
\end{figure}

Full modelling of the frequency conversion in the silicon nanowire is shown in Fig.~\ref{fig:silicon_sim}. The maximum conversion occurs for a pump power of 0.33~W, close to the analytic estimate. The maximum efficiency is 55\%, lower than the 85\% calculated for the desired conversion process alone. Most of the discrepancy is likely due to spurious BS upconversion. Cascaded pump MI and loss also contribute. Nevertheless, it is theoretically possible to achieve conversion efficiency above 50\%, suitable for HOM interference.

\section{Conclusion}

We have characterised the conditions for high efficiency and low-noise Bragg scattering frequency conversion in the single photon regime. The phase-matching conditions for the desired frequency conversion as well as spurious scattering and modulation instability were outlined. It was shown that third-order dispersion can suppress spurious processes while higher-order dispersion limits frequency conversion efficiency. The resulting optimisation procedure was applied to highly nonlinear fibre, silicon nitride waveguides and silicon nanowires. Numerical simulations confirmed that high single-photon frequency conversion efficiencies can be obtained. These design guidelines will support the development of efficient frequency conversion devices to interface together complex quantum networks and open new avenues to tunable, multicolor quantum processing.

\begin{acknowledgments}

This work was supported by the Centre of Excellence (CUDOS, project number CE110001018), Laureate Fellowship (L120100029) and Discovery Early Career Researcher Award programs (DE130101148) of the Australian Research Council (ARC).

We thank Chad Husko and Michael Steel for helpful discussions.
\end{acknowledgments}

\appendix

\section{Effective length approximation for propagation loss}
\label{sec:leff}

For parametric amplifiers with propagation loss, it is possible in the undepleted pump regime to obtain a closed-form solution in terms of confluent hypergeometric functions~\cite{marhic2008book}. However, for slowly varying properties, such a linear propagation loss, it is possible to obtain simpler solutions using the WKB approximation, also known as the phase integral method. For a parametric amplifier with total pump power $P_\text{tot}=P_1+P_2$ and number of pump $n_p$, the WKB method leads to a signal gain~\cite{marhic2008book}:
\begin{multline}
G_s(z) = e^{-\alpha z}\\
\times\bigg( 1 \pm e^{-\alpha z}\bigg| \frac{(2\gamma P_0)^2}{k(0)k(z)} \sin^2{\bigg( \int_0^zk(z')dz' 
\bigg)}  \bigg| \bigg).
\label{eq:gswkb}
\end{multline}
The parametric wavenumber $k(z)$ with loss is:
\begin{equation}
k(z) = \sqrt{(\kappa(z)/2)^2-i\partial_z\kappa(z)/2 \mp (2\gamma \sqrt{P_1P_2} e^{-\alpha z})^2} ,
\end{equation}
with the phase-mismatch $\kappa(z)$:
\begin{equation}
\kappa(z) = 2\delta - i\alpha + e^{-\alpha z}(3-n_p)\gamma (P_1\pm P_2).
\end{equation}
For BS we use the lower signs, while for MI we use the upper signs.

For BS we take $n_p=2$ and $P_0=P_1=P_2$, which yields the BS wavenumber:
\begin{equation}
k_{\text{bs}}(z) = \sqrt{(\delta_\text{bs} - i\alpha/2)^2 + (2\gamma P_0e^{-\alpha z})^2}.
\end{equation}
To first order, consider the case the BS process is centred at the zero-dispersion wavelength so that $\delta=0$. The wavenumber becomes:
\begin{equation}
k_{\text{bs}}(z) \approx 2\gamma P_0e^{-\alpha z}\sqrt{1-\bigg(\frac{\alpha e^{\alpha z}}{4\gamma P_0}\bigg)^2} \approx 2\gamma P_0e^{-\alpha z}.
\end{equation}
We have assumed that the loss term in the square root is small. In the case of silicon nitride and silicon modelled above, this is $\sim 0.06$ and can indeed be neglected.

The normalized power of the BS input signal then follows from Eq.~\ref{eq:gswkb}:
\begin{equation}
G_{\text{bs}}(L) \approx e^{-\alpha z}\big[ 1 +\sin^2{( 2\gamma P_0 L_{\text{eff}})}\big],
\end{equation}
where we have used $L_{\text{eff}}=(1-e^{-\alpha z})/\alpha$ from  Eq.~\ref{eq:pmax}. This is similar to the results leading to the power for maximum conversion, if we replace $L$ with $L_{\text{eff}}$ and reduce the conversion by the total propagation loss $e^{-\alpha z}$. A similar expression can be derived for MI around the optimal phase-matching condition $\kappa \approx 0$:
\begin{equation}
G_{\text{mi}}(L) \approx e^{-\alpha z}\big[ 1 +\sinh^2{( \gamma P_\text{tot} L_{\text{eff}})}\big].
\end{equation}

\bibliography{QuantumBragg}

\end{document}